\documentclass[twocolumn,aps,prl,superscriptaddress,unsortedaddress]{revtex4}
\newcommand{\etal}{{\it et al.}}

\begin{document}

\title{Comment on ``Circular Dichroism in the Angle-Resolved Photoemission Spectrum of
the High-Temperature Bi$_2$Sr$_2$CaCu$_2$O$_{8+\delta}$ Superconductor"}

\author{M. R. Norman}
\affiliation{Materials Science Division, Argonne National Laboratory, Argonne, IL 60439}
\author{A. Kaminski}
\affiliation{Ames Laboratory and Department of Physics and Astronomy,  Iowa State
University, Ames, IA  50011}
\author{S. Rosenkranz}
\affiliation{Materials Science Division, Argonne National Laboratory, Argonne, IL 60439}
\author{J. C. Campuzano}
\affiliation{Materials Science Division, Argonne National Laboratory, Argonne, IL 60439}
\affiliation{Department of Physics, University of Illinois at Chicago, Chicago, IL 60607}

\maketitle

In a recent Letter \cite{matti}, Arpiainen \etal~provide a structural explanation
for the dichroism observed in Ref.~\onlinecite{adam} by photoemission in the
pseudogap phase of underdoped Bi$_2$Sr$_2$CaCu$_2$O$_{8+\delta}$ (Bi2212).
Based on first-principles calculations, they claim that the observed dichroism is a
consequence of orthorhombic distortions.
We point out in this Comment that this scenario has a number of issues which
make it an unlikely source of the observed signal.

In their model, the origin of the calculated dichroism is a final state matrix element effect:
because the orthorhombic distortions are mostly confined to the BiO planes
and have very little effect on the CuO planes \cite{miles}, the electronic structure near $E_F$ is
almost unchanged, but the ARPES intensity is strongly altered by the distortion of the terminating
BiO layer. The attractive feature of this model is that it naturally explains the 
observed symmetry of the dichroism effect.  However,
thin films as used in Ref.~\onlinecite{adam} are heavily twinned, meaning any dichroism due to 
orthorhombic distortions should average to zero, and this feature was indeed one of the reasons to 
choose thin-film samples for the ARPES studies.

Second, as reported in Ref.~\onlinecite{adam}, 
the zero of the dichroism at high temperatures occurs at $M$ within experimental resolution.
In stark contrast to this, the calculation shows a large shift $\simeq 0.07 \pi/a_t$  (where $a_t$ is 
the tetragonal lattice constant) of the zero of the dichroism away from $M$, leading to a large 
dichroism signal at $M$ already at room temperature.  Arpiainen \etal~explain this discrepancy by
suggesting that the thin films may have a different structure, and that it is unclear what
photon energy to use to compare to experiment given uncertainties in the calculated final states.
However,
the main reason for the suppression of the superstructure intensity in ARPES and x-ray scattering, 
which is the sole indicator of a structural difference compared to bulk samples, is disorder.
And it would be a fortuitous accident if the zero is at $M$ for the experimental photon energy.

Moreover,  the temperature dependence of the signal calculated by Arpiainen \etal~varies strongly with 
momentum:  the dichroism is $T$ independent at 0.07 $\pi/a_t$ (e.g., the zero of the dichroism 
does not change with $T$), but the difference between low and high temperatures increases strongly with 
decreasing momentum (Ref.~1, Fig.~4). This is again in sharp contrast to the experimental data, 
where the variation of the dichroism signal with temperature is the same for all momentum points 
within $\pm$0.06 $\pi/a_t$, and also the zero of the dichroism moves away from $M$ as $T$ is 
lowered (Ref. 2, Figs. 3g, 4a, 4b).  Finally, we note that the structural
scenario does not give a natural explanation for the strong doping dependence
seen in Ref.~\onlinecite{adam}, in particular that the signal is confined to the pseudogap phase, and 
shows an order parameter-like temperature dependence below $T^*$.

The authors of Ref.~\onlinecite{matti} also comment on the null dichroism signal seen in Pb doped
samples in Ref.~\onlinecite{borisenko}, claiming that Pb doping might somehow restore the mirror
planes to their tetragonal locations.  We find this scenario rather difficult to believe as it is well known that Pb
doping leads to a {\it larger} orthorhombic deformation than seen in non Pb doped samples \cite{glady}.
And as with the thin films, these samples are also disordered.

In addition, the authors of Ref.~\onlinecite{matti} mention the x-ray dichroism study of
Kubota \etal~\cite{kubota}.  We would like to point out that the x-ray natural
circular dichroism signal of those authors (which matches the temperature dependence of
our signal) would be zero \cite{sergio} for the
centrosymmetric space group assumed in Ref.~\onlinecite{matti}.

On the other hand, we cannot rule out an orthorhombic distortion solely due to electronic degrees of
freedom (assuming that it does not domain average to zero). We note that a nematic effect has 
been recently seen in underdoped YBCO \cite{louis}.
But, in Bi2212, the orthorhombic axes are rotated 45 degrees relative to the tetragonal axes, so
a nematic deformation would be equivalent to a monoclinic distortion, which has not been
observed.

Work at Argonne National Laboratory and Ames Laboratory was supported by the U.S. DOE, Office of 
Science, under Contracts  No.~DE-AC02-06CH11357 and DE-AC02-07CH11358, respectively.

\end{document}